\documentstyle[12pt]{article}

\newcommand {\beq}{\begin{equation}}
\newcommand {\eeq}{\end{equation}}
\newcommand {\beqa}{\begin{eqnarray}}
\newcommand {\eeqa}{\end{eqnarray}}         
\newcommand {\beqs}{\begin{eqnarray*}}
\newcommand {\eeqs}{\end{eqnarray*}}
\newcommand {\bds}{\begin{displaymath}}
\newcommand {\eds}{\end{displaymath}}

\newcommand {\nn}{\nonumber}

\newcommand{\no}{\noindent} 

\newcommand {\bebb}{\begin{thebibliography}}
\newcommand {\eebb}{\end{thebibliography}}      
\newcommand {\bbit}{\bibitem}

\def\al{\alpha}

\def\ps{\psi}

\def\ot{\otimes} 
\def\op{\oplus}

\def\eqv{\equiv}

\def\journal#1&#2(#3){\unskip, \sl #1\ \bf #2 \rm(19#3) }
\def\andjournal#1&#2(#3){\sl #1~\bf #2 \rm (19#3) }

\def\np#1#2#3{Nucl. Phys. {\bf B#1}, #2(#3)}

\def\plb#1#2#3{Phys. Lett. {\bf B#1}, #2(#3)}

\def\cqg#1#2#3{Class. Quant. Grav. {\bf #1}, #2(#3)}

\begin{document}

\hfill{AMSS-1999-021}

\baselineskip = 18pt


\vskip 1cm

\begin{center}
{\LARGE\bf $W$-symmetries on the Homogeneous Space $G/U(1)^r$}

\vspace{1cm}

{\normalsize\bf
Xiang-Mao Ding $^{a,b}${\thanks {E-mail:xmding@itp.ac.cn; 
corresponding author}}, 
Pei Wang $^c$ {\thanks {E-mail:pwang@phy.nwu.edu.cn}}, 
}\\
\normalsize $^a$ CCAST, P.O. Box 8730, Beijing,100080, China\\
\normalsize $^b$ Institute of Applied Mathematics, \\
\normalsize Academy of Mathematics and Systems Science,\\
\normalsize Academia Sinica, \\
\normalsize P.O.Box 2734, Beijing, 100080, China\\
\normalsize  $^c$ Institute of Modern Physics, Northwest University, 
Xian, 710069, China\\
\end{center}

\date{}


\begin{abstract}

A construction of 
$W$-symmetries 
is given only in terms of the nonlocal 
fields (parafermions ${\ps}_{\al}$), 
which take values on the homogeneous space $G/U(1)^r$, 
where $G$ is a simply connected compact Lie group manifold (its 
accompanying Lie algebra ${\cal G}$ is a simple one of rank $r$).  
Only certain restriction of the root set of Lie algebra 
on which the parafermionic fields take values are satisfied, 
then a consistent and non-trivial extension of the stress 
momentum tensor may exist. 
For arbitrary simple-laced algebras, i.e. 
the $A-D-E$ cases, a more detailed discussion is given. 
The OPE of spin three primary field are calculated, 
in which a primary field with spin four is emerging.

\end{abstract}

\vspace{1cm}

\vspace{0.5cm}

 
\setcounter{section}{1}
\setcounter{equation}{0}
\section*{1.Introduction}

The discovery of the $AdS_{d+1}/CFT_d$ correspondence \cite{Mad,GKP,Witt} 
brings the role of conformal field theory to a special stage. The type 
$II$B string theory on the $AdS_{5}\ot S^5$ are equivalent to ${\cal N}
= 4$ super Yang-Mills in Minkowski space-time. However, the calculation 
of the correlation functions of physical quantities are limited by our 
knowledge, except for the $CFT_2$, saying 2d conformal field theory (CFT) 
case. With the help of infinitely dimensional symmetries of 2d CFT, 
much more information can be obtained. For example, 
Seiberg et al. considered the duality between 
$AdS_3$ and $CFT_2$ \cite{GKS,KS}. 
As a special case of this duality, a type $II$B string theory on 
$AdS_3\ot S^3\ot T^4$ is equivalent to a certain 2d superconformal 
field theory (SCFT), which corresponds to the IR 
limit of the dynamics of parallel $D1$-branes and $D5$-branes. 
Light-cone gauge quantization of string theories on $AdS_3$ are 
given in \cite{YZh}. 

For 2d CFT, when the central charge of the theory is greater than one, 
the Virasoro symmetry must be enlarged \cite{Zamo}, or more primary 
fields should be added, this extended structure 
is called $W$ symmetry. If there are a finite number of primary fields, 
then the values of central charge $c$ and conformal weight ( or spin) 
$h$ take on rational values. there are called rational conformal field 
theories (RCFT), see \cite{BoSc,FORT,Hull,West} for review. 

The $Z_k$ parafermion (PF) algebra is proposed by Zamolodchikov 
and Fateev \cite{ZaFa} for describing a two-dimensional 
statistical system with $Z_k$ symmetry 
associating "spin" variables $\sigma _r$ to each node $r$ in a 
(square) lattice $L$, the $\sigma _r$ take the $N$ values 
$\omega ^{q}\;(q=0,\;1\;,\ldots,
k-1)$, where $\omega =exp(2i\pi/k)$. This generalizes 
the fermion of the Ising model, which corresponds to the node 
of $Z_2$. It is also known that there are various of statistical 
models, which can be  described by this 
extended theory, such as the $3$-state Potts model 
($k=3$) \cite{FaZa}, Ashkin-Teller model ($k=4$) \cite{Yang}.  
The $Z_4$ parafermion also gives a consistent 6d string 
theory \cite{Arg}. 

In fact, parafermion field is important in fractional 
superstring theory \cite{CES}, $W$-string 
theory \cite{West,Hull}, furthermore in the compactification 
of a type $II$ string theories on a Calabi-Yau (CY) 
manifolds \cite{Gepn,Gep} and the construction of 
${\cal N}=2$ SCFT \cite{DiQi,GeQi}. Gepner model is the 
tensor products of ${\cal N}=2$ minimal models with the 
internal central charge $c=9$, which is exactly a solvable 
models for strings compactified on a CY manifolds. 
And its applications in $Dp$-brane theory are 
also presented in \cite{RSc,GSa} recently. 
For example, $D0$-branes, the wrapping of $Dp$-branes on 
$p$-dimensional supersymmetric cycles leads to BPS saturated. 
Its dynamics can be analyzed by a Ishibashi boundary 
states \cite{Ish}. The Ishibashi boundary state is the RCFT 
extension of a boundary 
state of open sting theories. In open sting theories, 
the boundary must be chosen such that the 2d CFT symmetry is 
not broken \cite{Card,Green,PCai}. 

\beq
(L_n -{\bar L}_{-n})|B>=0,
\eeq

\no here $|B>$ is a boundary state. When the extension structure 
of the CFT forms a RCFT, the RCFT symmetry on the boundary 
must be hold also. So that the bulk left- and right-moving 
primary currents $W$, ${\bar W}$ have to satisfy certain 
relations on the boundary. On the construction of the boundary 
state, the Ishibashi states $|i>>$ have the following relation, 
   
\beq
(W_n -(-1)^{h_W}{\bar W}_{-n})|i>>=0,
\eeq

\no where the $h_W$ is the conformal dimension of $W$. 
So the explicit expression of $W$ algebra is important and helpful 
for solving this problem. For the intrinsic relations between the Gepner 
model and the parafermion \cite{Gep}, the construction of $W$ algebra 
without introducing the free boson has his own advantage.
 
It is well known that conformal algebra may be obtained from  
current algebra via Sugawara construction. Similar ways of 
constructing $W$-algebra from current algebras  
were found by Bais et al. \cite{BBSS}, through simple third order 
Casimir in level one case and $GKO$ coset model in 
$SU(N)_1\otimes SU(N)_k/SU(N)_{k+1}$ case. The construction 
of $W$-algebra directly (not by free field realization ) 
from $SU(2)_k$ parafermion was also proposed \cite{WaDi}, 
in which the so called $Z$-algebra technique was used. 
In a sense of that the generating PFs 
can be defined through the current algebras by projecting out the 
Cartan subalgebraic valued components, the $Z$-algebra construction 
may have the most similarity to the Sugawara construction. On the 
other hand, parafermion is a coset valued field. Thus the 
parafermion realization of $W_k$ algebra for specific level 
may unify the Sugawara and the coset construction. 
It has been conjectured that all RCFT can be represented as 
cosets, and that any CFT can be arbitrary well approximated by 
a rational theory. So the studying of rational theory has his 
own interesting. As we know, the bosonization representation of 
a conformal model provide a much bigger Fock 
space\cite{DiQi,Neme,Waki}. We have to use 
the BRST operator ($Q_{BRST} ^2=0$ or other restrictions) 
to project it onto the physical 
space. Hence there are much more complications if we use the 
free field realization. Therefore the advantage of 
our approach is that it avoids the ambiguity 
and complexity of the bosonization. It is also worthy to find $W_n$ 
algebras for PFs of higher rank group. The reason is the follows. 
The central charges of $PFs$ \cite{Gepn,GeQi,ZaFa} 

\begin{equation}
c=\frac{kD}{k+g}-r,
\label{eq:c}
\end{equation}

\noindent where $D$, $r$ are dimension, rank of Lie algebra ${\cal G}$ 
respectively, and level $k$ of ${\hat {\cal G}}$ is also an integer 
defining the cyclic symmetry of PFs. In $SU(2)_k$ case 
$c=2(k-1)/(k+2)$ agrees with a special case in the Fateev-Lykyanov's 
$W$-algebra series $ c=(k-1)[1-k(k+1)/p(p+1)]_{p=k+1}$ \cite{FaLy}. 
However, there is no known $W$-algebra, which is constructed 
from boson, current algebra or coset model, has the central 
charge coinciding with the PFs construction of groups with 
higher rank. 

In \cite{DFSW,DFSh,WaDi} we gave a construction of Virasoro 
algebra by using non-local fields (parafermions) which take 
values on coset space $G/U(1)^r$, where $G$ is a simply connected 
compact Lie group manifold, its Lie algebra ${\cal G}$ 
is a simple one with rank $r$. There the so called $Z$-algebra 
technique was used. We also extended this approach to construct 
the $W$-symmetries, $W_3$ algebra and $W_5$ algebra were 
obtained from $SU(2)$ and $SU(3)$, respectively (part results 
in $SU(2)$ case was reconsidered recently in \cite{Maro}). 
While in ref. \cite{DFSW,DFSh} the construction of 
$W_3$ algebra from the $SU(3)$ parafermion was based on a 
special choice of the root set for summation, and turned out 
that the $W$ algebra were magical closed, while for other choice 
of root set, the construction was not correct. 

As known that in ref. \cite{DFSW,DFSh,WaDi} we only obtained 
the PFs construction of RCFT for $SU(n)(n\leq 3)$ cases, 
and further extension of this construction was not succeed. 
In fact, from $SU(4)$ PFs the next and direct goal this 
construction is failed for spin three primary field. However, 
it seems that the possibility was not removed at any extent, 
and the reasons what make these problems arising were unclear 
at that time. In this paper we will discuss these problems. 

The layout of this paper is as follows. In section $2$, we recall 
some basic aspects of extended CFT and the parafermion field, 
establish our notations and obtain the identities which will be 
used at a later stage. In section $3$, using the relations 
obtained in section $2$, the Virasoro algebra constructed from 
arbitrary Lie algebra ${\cal G}$ PFs is given, the approach 
presented here greatly simplifies the calculation in 
\cite{DFSW,DFSh,WaDi}. If we hope that  the extension structure 
of RCFT is nontrivial, certain restriction must be put for the 
Lie algebra root set $\Phi $ on which the parafermion 
fields take values. They coincide with the known results for the 
$SU(2)$ and $SU(3)$ cases, and get rid of the possibility by this 
construction from $SU(4)$ PFs for spin three primary field. 
Assuming the condition of the root set is satisfied, in section $4$ 
we obtain a spin $3$ primary field very general in simple-laced 
case, more detailed discussion is given for $A_l$ algebra case.

\vspace{1cm}

\setcounter{section}{2}
\setcounter{equation}{0}
\section*{2.Brief review of RCFT}

In this section we first review some basic aspects of Virasoro algebra, 
$W$ algebra and parafermion field. Then notations which will be 
used in the sequel are introduced.

The OPE of the stress momentum tensor is  

\begin{equation}
T(z)T(w)=\frac{c/2}{(z-w)^4}+\frac{2T(w)}{(z-w)^2}+
\frac{\partial T(w)}{z-w}+\ldots,
\end{equation}

\no the commutator of its modes generate a chiral algebra 
which is just the Virasoro algebra. To simplify the the 
expression of OPE, we denote the last equation as,
 
\begin{equation}
 [TT]_ 4=c/2,\hskip 0.5truecm  [TT]_ 3=0,\hskip 0.5truecm
 [TT]_ 2=2T,\hskip 0.5truecm  [TT]_ 2=\partial T,\hskip 0.5truecm.
\end{equation}

As mentioned previously in the introduction, in the case of the 
central charge greater than one, it is necessary to enlarge the 
symmetry of the CFT by adding primary field with spin 
great than two \cite{Zamo}. The first nontrivial chiral primary 
field $W(z)$ of conformal dimension $3$ (For the left chiral field 
its conformal dimension is identical with its spin, so we use 
them without difference.) with the OPE,
  
\begin{eqnarray}
W(z)W(w)=& &\frac{c/3}{(z-w)^6}+\frac{2T(w)}{(z-w)^4}+
\frac{\partial T(w)}{(z-w)^3}\nonumber\\
& &+\frac{1}{(z-w)^2}(2b^2\Lambda (w)+\frac{3}{10}\partial ^2T)\nonumber\\
& &+\frac{1}{z-w}(b^2\partial\Lambda (w)+\frac{1}{15}\partial ^3T)+\ldots,\\
\label{eq:W3}
T(z)W(w)=& &\frac{3W(w)}{(z-w)^2}+\frac{\partial W(w)}{z-w}+\ldots,
\label{eq:TW3}
\end{eqnarray}

\noindent where $\Lambda (z)=[TT]_0(z) -\frac{3}{10} \partial ^2 T(z)$, 
and constant $b^2=16/(22+5c)$. The primary feature of $W$ field is 
governed by the equation (\ref{eq:TW3}). Identically, we express 
the above equation as 

\begin{displaymath}
 [WW]_ 6=c/3,\hskip 0.3truecm  [WW]_ 5=0,\hskip 0.3truecm
 [WW]_ 4=2T,\hskip 0.3truecm   [WW]_ 3=\partial T, 
\end{displaymath}
\begin{equation}
 [WW]_2=(2b^2\Lambda (w)+\frac{3}{10}\partial ^2T),\hskip 0.3truecm 
 [WW]_ 1=(b^2\partial\Lambda (w)+\frac{1}{15}\partial ^3T),
\end{equation}
\begin{equation}
 [TW]_{6,5,4,3}=0,\hskip 0.3truecm  [TW]_2=3W,\hskip 0.3truecm 
 [TW]_1=\partial W.
\end{equation}

Parafermionic currents are primary fields of the 2d CFT. The general 
parafermion defined for root lattices are proposed in \cite{Gepn}. 
For a semi-simple Lie algebra ${\cal G}$ there are $D-r$ generating 
parafermion operators $\psi _{\alpha}$, where $D=dim {\cal G}$ 
and $r=rank {\cal G}$ are dimension and rank of ${\cal G}$, 
respectively. $\alpha$ is a root of ${\cal G}$ (analogy to their 
counter part in the antiholomorphic sector $\bar{\psi} _{\alpha}$, 
which will be left out for simplicity). For general parafermion 
we denote them by a vector in the root lattices $M$ mod a 
lattice $kML$, where $ML$ is the long root lattices and $k$ 
is a constant identified with the level in the corresponding affine Lie 
algebra $\widehat{\cal G}$ \cite{Kac}. The generating parafermions are 
defined through their relationship with current algebra. Thus define the 
fields  

\begin{eqnarray}
\chi _{\alpha}(z)=\sqrt{\frac{2k}{\alpha ^2}}:\psi _{\alpha}(z)
exp(i\alpha\phi(z)/\sqrt{k}):,\nonumber\\
h_j(z)\equiv h_{\alpha _j}(z)=\frac{2i\sqrt{k}}{\alpha ^2 _j} {\alpha}_j
\partial _z \phi (z),
\end{eqnarray}

\noindent for any root $\alpha$ and simple root $\alpha _j$. We  
require that the currents $\chi (z)$ and $h(z)$ (Cartan subgroup 
valued components) obey the OPE of the current algebra. 
Because of the mutually semi-local property between the two 
parafermions, the radial ordering product is a multivalued functions, 
so we can define the radial order product of (generating) parafermions 
(PFs) $\psi _{\alpha}(z), \;\psi _{\beta}(w)$ ($\alpha, \;\beta$ are roots 
of the underlying Lie group)\cite{DFSW}

\begin{eqnarray}
&&R(\psi _{\alpha}(z)\psi_{\beta}(w))\nonumber\\
&&=\left\{
\begin{array}{ll}
\psi _{\alpha}(z)\psi_{\beta} (w),
\;\ldots, \omega ^{(k-1)\alpha\beta}\psi _{\alpha}(z)\psi _{\beta}(w),
\hskip 0.2truecm |z|>|w|\;;\\
\omega ^{\alpha\beta/2}\psi _{\beta}(w)\psi _{\alpha}(z),\;\ldots
\omega ^{(k-1/2)\alpha\beta}\psi _{\beta}(w)\psi _{\alpha}(z),
\hskip 0.2truecm |z|<|w|\;,
\end{array}
\right.
\label{eq:radp}
\end{eqnarray}

\noindent where $\omega =exp(2\pi i/k)$. The RHS of (\ref{eq:radp}) 
is the requirement of analysis for the field. By using 
(\ref{eq:radp}) one therefore have the following relation for the 
parafermion fields  

\begin{equation}
R\left(\psi _{\alpha}(z)\psi _{\beta}(w)\right)(z-w)^{\alpha\beta/k}
=R\left(\psi _{\beta}(w)\psi _{\alpha}(z)\right))(w-z)^{\alpha\beta/k}.
\label{eq:rlrr}
\end{equation}

\noindent which is an extension of that for fermion (i.e. 
$\alpha\cdot\beta=1,\;k=2$), and boson (i.e. $k\rightarrow \infty $). 
We will drop the $R$ symbol in the following without confusion. 
The OPE of the parafermion fields defined by\cite{DFSW}

\beqa
\psi _{\alpha}(z)\psi _{\beta}(w)(z-w)^{\alpha\beta/k}
&&=\frac{\delta _{\alpha,-\beta}}{(z-w)^2}
+\frac{\varepsilon _{\alpha,\beta}/{\sqrt k}}{z-w}
\psi _{\alpha +\beta}(w)\nn\\
& &~~~+ \sum _{n=0} ^{\infty} (z-w)^n 
[ \psi _{\alpha} \psi _{\beta} ]_{-n},\nn\\
&&\equiv \sum _{n=-2} ^{\infty}(z-w)^n [\psi _{\alpha}\psi _{\beta}]_{-n},
\label{eq:Par}
\eeqa

\noindent Which means that we have 

\begin{equation}
 [ \psi _{\alpha}\psi _{\beta} ] _l=0,\;(l\geq 3)\hskip 0.3truecm
 [ \psi _{\alpha}\psi _{\beta} ]_2=\delta _{\alpha,-\beta},
\hskip 0.3truecm 
 [\psi _{\alpha}\psi _{\beta} ]_1=\frac
 {\varepsilon _{\alpha,\beta}}{\sqrt k}\psi _{\alpha +\beta}
\end{equation}

\noindent where $\varepsilon _{\alpha,\beta}$ is the structure constant of 
Lie algebra ${\cal G}$ (see more details in the Appendix). 

For every field in the parafermion theory there is a pair of charges 
$(\lambda, \bar{\lambda})$, which take values in the weight lattice. 
So we denote such field by $\phi _{\lambda, \bar{\lambda}}(z,\bar{z})$ 
\cite{ZaFa,Gepn,DFSW}. The OPE of the $\psi _{\alpha}$ and 
$\phi _{\lambda, \bar{\lambda}}(z,\bar{z})$ is given by 
 
\begin{equation}
\psi _{\alpha}(z)\phi _{\lambda, \bar{\lambda}}(w,\bar{w})
=\sum _{-\infty} ^{\infty}(z-w)^{-m-1-\alpha\lambda}
A_m ^{\alpha,\lambda}
\phi _{\lambda, \bar{\lambda}}(w,\bar{w})
\end{equation}

\noindent which means that we define the action of the operator (mode)
$A_m ^{\alpha,\lambda}$ on $\phi _{\lambda, \bar{\lambda}}(z)$ by 
the integration

\begin{equation}
A_m ^{\alpha,\lambda}\phi _{\lambda, \bar{\lambda}}(w,\bar{w})
=\oint _{c_w}\;dz\;(z-w)^{m+\alpha\lambda}
\psi _{\alpha}(z)\phi _{\lambda, \bar{\lambda}}(w,\bar{w})
\end{equation}

\noindent where $c_w$ is the contour around $w$, and for simplicity 
the notation $\oint \;dz \equiv \oint\; \frac{dz}{2\pi i}$ is implied.    

Assuming that fields $A_{\alpha}$ and $B_{\beta}$ are  
arbitrary function of parafermions with parafermion charges $\alpha$ 
and $\beta$. The fields are local ($\alpha$,or $\beta=0$) or semilocal 
($\alpha= \beta$=root of the underlying Lie algebra). The OPE of them 
can be written as 

\begin{equation}
R(A_{\alpha}(z)B_{\beta}(w))(z-w)^{\alpha\beta/k}=
\sum_{n=- [h_A+h_B ] }^{\infty} [AB ]_{-n}(w)(z-w)^n,
\end{equation}  

\noindent in which $ [ h_A ] $ means the integral part of dimension $A$. 
Hence we have

\begin{equation}
 [ A_{\alpha}(z)B_{\beta}]_n(w))=\oint_w \ dz \ A_{\alpha}(z)
B_{\beta}(w)(z-w)^{n-1+\alpha\beta/k}
\end{equation}  

\noindent and some relations

\begin{equation}
 [ \partial A_{\alpha}(z)B_{\beta} ]_{-n}(w)=(n+1-\alpha\beta/k) 
 [ A_{\alpha}(z)B_{\beta} ]_{-(n+1)}(w)
\end{equation}

\begin{equation}
 [A_{\alpha}\partial B_{\beta} ]_n(w)=(n-1+\alpha\beta/k)
 [ A_{\alpha}(z)B_{\beta} ]_{n-1}(w)
+\partial  [ A_{\alpha}(z)B_{\beta} ]_n(w)
\end{equation}

\begin{equation}
 [ \partial A_{\alpha}(z)B_{\beta} ]_{-n}(w)
 +[A_{\alpha}\partial B_{\beta} ]_{-n}(w)
=\partial  [ A_{\alpha}(z)B_{\beta} ]_{-n}(w)
\end{equation}

\begin{eqnarray}
 [ \partial ^n A_{\alpha}(z)B_{\beta} ]_0(w)
=(n-\alpha\beta/k)\ldots (1-\alpha\beta/k) 
 [ A_{\alpha}(z)B_{\beta} ]_{-n}(w)\nonumber\\
\equiv \frac{\Gamma (n+1-\alpha\beta/k)}{\Gamma(1-\alpha\beta/k)} 
 [ A_{\alpha}(z)B_{\beta} ]_{-n}(w)
\end{eqnarray}

\noindent in which the $\Gamma$ is the usual $\Gamma$- unction. 
It is easy to find a relation between three-fold radial 
ordering products

\begin{eqnarray}
& &\left\{\oint_w \ du \oint_w \ dz \ R(A(u)R(B(z)C(w)))\right.\nn\\
& &-\oint_w \ dz \oint_w \ du \ (-)^{\alpha\beta/k}R(B(z)R(A(u)C(w)))\nn\\
& &-\left. \oint_w \ dz \oint_z \ du \ R(R(A(u)B(z))C(w))\right\}\nn\\
& &(z-w)^{p-1+\beta\gamma/k}(u-w)^{q-1+\gamma\alpha/k}
(u-z)^{r-1+\alpha\beta/k}=0,  
\label{eq:threef}
\end{eqnarray}

\noindent where the integers $p,\;q,\;r$ are in the region $-\infty< \,p\, 
\leq  [h_B+h_C ] ,\;
\;-\infty<\,q\,\leq \, [h_C+h_A ] ,\;\;-\infty<\,r\,\leq\, [h_A+h_B ] $, 
and $\alpha,\;\beta,\;\gamma$ are parafermionic charges of the fields 
$A,\;B,\;$ and $C$ respectively. 
This equation is an extension of the identity $A(BC)-B(AC) - [A,B]C=0$. 
The contours are self evident. Performing the binomial expansion, we 
can rewrite the last equation as

\begin{eqnarray}
& & \oint_w \ du \oint_w \ dz \ R(A(u)R(B(z)C(w)))
\sum _{i=p} ^{\infty}C_{r-1+\alpha\beta/k} ^{(i-p)}\nonumber\\
& &~~~\times (z-w)^{i-1+\beta\gamma/k}(u-w)^{Q-1+(\beta+\gamma)\alpha/k}\nonumber\\
& &~~~+(-1)^r\oint_w \ dz \oint_w \ du \ R(B(u)R(A(z)C(w)))
\sum _{j=q} ^{\infty}C_{r-1+\alpha\beta/k} ^{(j-q)} \nonumber\\
& &~~~\times (z-w)^{Q-j-1+\beta(\alpha+\gamma)/k}(u-w)^{j-1+\gamma)\alpha/k}\nonumber\\
& &= \oint_w \ dz \oint_z \ du \ R(R(A(u)B(z))C(w))
\sum _{l=r} ^{\infty}C_{q-1+\alpha\gamma/k} ^{(l-r)}\nonumber\\
& &~~~\times (z-w)^{Q-l+(\alpha+\beta)\gamma/k}(u-z)^{l-r+\beta\alpha/k},
\label{eq:Bin}
\end{eqnarray}

\noindent From the two equations above we obtain the 
following Jacobi-like identity relations\cite{DFSW,WaDi} 

\begin{eqnarray}
&&\sum_{i=p}^{ [h_B+h_C ] }C_{r-1+\alpha\beta/k}^{(i-p)}
[A [BC ]_i ]_{Q-i}(w)\nonumber\\
&&~~~+(-)^r\sum_{j=q}^{ [h_C+h_A ] }C_{r-1+\alpha\beta/k}^{(j-q)}
[B [AC ]_j ]_{Q-j}(w)\nonumber\\
&&=\sum_{k=r}^{ [h_B+h_A ] }(-)^{(k-r)}C_{q-1+\alpha\gamma/k}^{(k-r)}
[ [AB ]_kC ]_{Q-k}(w),
\label{eq:Jacobi}
\end{eqnarray}

\noindent in which $ Q=p+q+r-1,C_x^{(l)}=\frac{(-)^lx(x-1)...(x-l+1)}{l!}$, 
and $C_0^{(0)}=C_n^{(0)}=C_{-1}^{(l)}=1,\;C_p^{(l)}=0$, for $p,l>0, l>p$. 
This identity is important for our usage, we will use it extensively.
Performing analytic continuation one more equation is obtained

\begin{equation}
 [ BA ]_r(w)=\sum_{i=r}^{ [h_A+h_B ] }\frac{(-)^t}{(t-r)!}
\partial^{t-r} [AB ]_t(w),
\label{eq:analytic}
\end{equation}

\noindent and two special cases should be mentioned $(n\geq 0)$

\begin{equation}
 [ A const. ]_{-n}(w)=const.\frac{1}{n!}\partial^n A(w),\hskip 0.5truecm
 [ const.A ]_{-n}(w)=const.\delta_{n,0}B(w).
\end{equation}

\noindent In all of the previous equations $A,\;B,\;C$ can be compound 
operators. We can calculate any coefficient in OPE from fundamental 
equation (\ref{eq:Par}).

\setcounter{section}{3}
\setcounter{equation}{0}
\section*{3.PFs constructions CFT}

In this section, we present an another approach beside the 
$Z$-algebra technique used in\cite{DFSW}. This approach greatly 
simplifies the calculation of \cite{DFSW}, and the restriction 
on the root set arises naturally from the definition of the 
primary parafermion field. The detailed derivation of OPE of 
the stress momentum tensor is given.

For the notation conveniences, define $N_{\alpha,\beta} 
\eqv\varepsilon _{\alpha,\beta}/{\sqrt k}$, 
and the following identities are hold by $N_{\alpha, \beta}$, 

\begin{equation}
N_{\alpha,\beta}=
-N_{\beta,\al}=-N_{-\alpha,-\beta}
=\frac{(\alpha+\beta)^2}{\beta ^2}N_{-\alpha,\alpha+\beta}.
\label{eq:N}
\end{equation}

\no If we only consider the simple-laced case, the results are 

\begin{equation}
N_{\alpha,\beta}=-N_{\beta,\al}=-N_{-\alpha,-\beta}
=N_{-\alpha,\alpha+\beta}. 
\end{equation}

\noindent Further more, we define 
$\tau _{\alpha} =[\psi _{\alpha}\psi_{-\alpha}]_0$, 
$\eta _{\alpha} =[\psi _{\alpha}\psi_{-\alpha}]_{-1}$, 
$\Omega _{\alpha} =[\psi _{\alpha}\psi_{-\alpha}]_{-2}$.
We calculate the OPE of $\tau _{\alpha}$ with $\psi _{\alpha}$, and 
$\tau _{\beta}$. The results coincide with the OPE of stress momentum 
tensor, and the modes of $\tau _{\beta}$ give the Virasoro algebras. 

From the definition of $\tau _{\beta}$, and the (\ref{eq:Jacobi}), 
obviously we have,
 
\begin{eqnarray}
\tau _{\alpha}=\tau _{-\alpha},\\
\left[\tau _{\alpha} \psi _{\beta} \right ]_l=0,(l\geq 3),
\end{eqnarray}

\noindent setting $Q=p=2,\;q=1,\;r=0$ in the (\ref{eq:Jacobi}), we have 

\begin{eqnarray}
& & [\tau _{\alpha} \psi _{\beta}]_2\\
& &=[\tau _{-\alpha} \psi _{\beta}]_2\nonumber\\
& &=[\psi _{\alpha}[\psi_{-\alpha}\psi _{\beta}]_2]_0
+[\psi _{-\alpha}[\psi_{\alpha}\psi _{\beta}]_1]_1 
+(1+{\alpha}^2/k)[\psi _{-\alpha}[\psi_{\alpha}\psi _{\beta}]_2]_0\nonumber\\
& &~~~-\frac{\alpha\beta}{k}[[\psi _{\alpha}\psi _{-\alpha}]_1 
\psi _{\beta}]_1 
+\frac{\alpha\beta}{2k}
\left(1-\frac{\alpha\beta}{k}\right)[[\psi _{\alpha} 
\psi_{-\alpha}]_2 \psi _{\beta}]_0\nonumber\\
& &=\delta _{\alpha,\beta}\psi _{\alpha}
+N_{\alpha, \beta}N_{-\alpha,\alpha+\beta}\psi _{\beta}
+(1+{\alpha}^2/k)\delta _{\alpha,\beta}\psi _{-\alpha}\nn\\
& &~~~+\frac{\alpha\beta}{2k}(1-\frac{\alpha\beta}{k})\psi _{\beta}\nn\\
& &=\delta _{-\alpha,\beta}\psi _{-\alpha}
+N_{-\alpha, \beta}N_{\alpha,-\alpha+\beta}\psi _{\beta}\nn\\
& &~~~+(1+{\alpha}^2/k)\delta _{-\alpha,\beta}\psi _{\alpha}
-\frac{\alpha\beta}{2k}(1+\frac{\alpha\beta}{k})\psi _{\beta}.
\label{eq:tap2}
\end{eqnarray}

\noindent We denote $\tau= \sum _{\alpha \in \Phi} \tau _{\alpha}$, 
where the $\Phi$ is the root set for summation. We require that 
the set satisfy the conditions 

\begin{displaymath}
\{\Phi\}\bigcap \{-\Phi\}=\emptyset,\; \{\Phi\}\bigcup \{-\Phi\}=\Delta, 
\end{displaymath}
\noindent Obviously the number of roots in $\Phi$ equals the number  
of ones in $P$. In fact, the $\Phi$ can be obtained from $P$ by 
some Weyl reflections. From the equation (\ref{eq:tap2}) we obtain 

\begin{equation}
 [\tau \psi _{\beta}]_2
=\left(1+\sum _{\alpha \in \Phi} (\alpha\beta/2k
-(\alpha\beta)^2/2k^2+N_{\alpha, \beta}N_{-\alpha,\alpha+\beta})\right)
\psi _{\beta},\; (\beta \in\Phi),
\label{eq:stap2}
\end{equation}

\noindent where without loosing generality we choose 
$\beta \in\Phi$ for convenience, and we will not mention it in the 
later stage. From the general theory of the conformal fields 
\cite{BPZ,Zamo}, we know that the conformal dimension of the parafermion 
$\psi _{\alpha}$ is $(1-\alpha ^2/2k)$. We normalize the $\tau$ to   

\begin{equation}
T=\frac{k}{k+g}\tau,
\end{equation}

\noindent in which the $g$ is the dual Coxeter number, and $ k$ is 
the level of the representation of 
$\hat{\cal G}$, which is the affinization of the classical Lie algebra 
${\cal G}$. For a consistent theory, we require 

\begin{equation}
 [T \psi _{\beta}]_2=\left(1-\frac{\beta ^2}{2k}\right)\psi _{\beta}
\label{eq:TP2}
\end{equation}

\noindent or, equivalently,

\begin{equation}
 [\tau \psi _{\beta}]_2
=\left(1+\frac{2g-\beta ^2}{2k}-\frac{g\beta ^2}{2k^2}\right)\psi _{\beta}
\label{eq:ctap2}
\end{equation}

\noindent Comparing the last equation with (\ref{eq:stap2}), we get the 
following conditions for set $\Phi$.

\begin{equation}
\sum _{\alpha \in \Phi}\left(\frac{\alpha\beta}{2k}
+N_{\alpha, \beta}N_{-\alpha,\alpha+\beta}\right)
=\frac{2g-\beta ^2}{2k},
\label{eq:con1}
\end{equation}
\begin{equation}
\sum _{\alpha \in \Phi} (\alpha\beta)^2=
\sum _{\alpha \in P} (\alpha\beta)^2=g \beta ^2
\label{eq:con2}.
\end{equation}

\noindent The last two equations are just the consistent 
condition for PFs construction of CFT. 
From the definition of $g$ we know that the condition 
(\ref{eq:con2}) is satisfied for any given Lie algebra 
($\psi ^2=2$). While the condition (\ref{eq:con1}) brings a 
constraint on root system of ${\cal G}$. 
Therefore we get a necessary condition for the root set 
on which the summation is defined for a consistent theory. 
From (\ref{eq:tap2}) we obtain: 

\begin{equation}
k\sum _{\alpha \in \Phi}\left(N_{\alpha, \beta}N_{-\alpha,\alpha+\beta}
-N_{-\alpha, \beta}N_{\alpha,-\alpha+\beta}\right)=\beta ^2,
\end{equation}

\noindent while on the other hand, we have  

\begin{equation}
k\sum _{\alpha \in \Phi}\left(N_{\alpha, \beta}N_{-\alpha,\alpha+\beta}
+N_{-\alpha, \beta}N_{\alpha,-\alpha+\beta}\right)=2g-2\beta ^2.
\end{equation}

\noindent So we get the solution

\begin{equation}
k\sum _{\alpha \in \Phi}N_{\alpha, \beta}N_{-\alpha,\alpha+\beta}
=\frac{2g-\beta ^2}{2},
\end{equation}

\begin{equation}
k\sum _{\alpha \in \Phi}N_{\alpha,-\beta}N_{-\alpha,\alpha-\beta}
=\frac{2g-3\beta ^2}{2},
\end{equation}

\noindent and we have (if $N_{\alpha ,\beta} \neq 0 $, or,
$N_{\alpha,-\beta} \neq 0$)

\begin{equation}
\sum _{\alpha \in \Phi}(\alpha\beta)=0,
\label{eq:cph1}
\end{equation}

\noindent in which $\beta$ is an arbitrary element of the $\Delta$, 
so we can re-express the last identity as, 

\begin{equation}
\sum _{\alpha \in \Phi}\alpha =0,
\label{eq:cph2}
\end{equation} 

\noindent This is the condition for root system ${\cal G}$ 
on which the summation will be taken over. Which says that 
for very simple root $\al _i$ the sum of his height in $\Phi$ 
must be zero.  For $SU(3)_k$ as an example 
$\Phi =\{\al _1, ~\al _2, ~\al _3 =-(\al _1 + \al _2, ) \}$, this 
coincides with the result in \cite{DFSW}. In this paper we only 
consider the simple-laced cases for simplicity, 
saying $\alpha ^2=2$, and $\alpha\beta=-1$, if $\alpha+\beta \in\Delta$. 
When $g=2, \;N_{\alpha, \beta}=0$, this is a special case, 
and we have $\alpha=\beta$, this is the $SU(2)_k$ ($Z_k$ symmetry).  
please see \cite{WaDi} for more details.
While for $g\geq 2$, we have the following identities:  

\begin{equation}
k\sum _{\alpha \in \Phi}N_{\alpha, \beta}N_{-\alpha,\alpha+\beta}=g-1,\;\;
k\sum _{\alpha \in \Phi}N_{-\alpha, \beta}N_{\alpha,-\alpha+\beta}=g-3,
\end{equation}

\begin{equation}
k\sum _{\alpha \in \Phi}\alpha\beta 
N_{\alpha, \beta}N_{-\alpha,\alpha+\beta}=-(g-1),\;\;
k\sum _{\alpha \in \Phi}\alpha\beta
N_{-\alpha, \beta}N_{\alpha,-\alpha+\beta}=g-3,
\end{equation}
\begin{equation}
\sum _{\alpha \in \Phi}(\alpha \beta)^2=2g,\hskip 0.5truecm
\sum _{\alpha \in \Phi}(\alpha \beta)^3=6,\hskip 0.5truecm
\sum _{\alpha \in \Phi}(\alpha \beta)^4=2g+12.
\end{equation}

\no The proof of the above identities is very simple. 
Using these identity (we will not mention them separately), 
we have

\begin{equation}
 [T\psi _{\beta}]_2=(1-1/k)\psi _{\beta}
\end{equation}

\noindent repeating the same procedure, we have

\begin{equation}
 [T\psi _{\beta}]_1=\partial \psi _{\beta}
\end{equation}

\noindent in the process of deriving the last equation, the identity,  

\begin{eqnarray}
\sum _{\alpha \in \Phi}(N_{\alpha, \beta}[\psi _{-\alpha}
\psi _{\alpha+\beta}]_0
+N_{-\alpha, \beta}[\psi _{\alpha}\psi _{-\alpha+\beta}]_0)\nonumber\\
=\frac{1}{2} \sum _{\alpha \in \Phi}
\left( N_{\alpha, \beta}N_{-\alpha,\alpha+\beta}
+N_{-\alpha, \beta}N_{\alpha,-\alpha+\beta} \right)\partial \psi _{\beta},
\end{eqnarray}

\noindent is used. We can express the results as the OPE

\begin{equation}
 T(z)\psi _{\beta}(w)=\frac{1-1/k}{(z-w)^2}
+\frac{1}{z-w}\partial \psi _{\beta}(w)+\ldots.
\end{equation}

\noindent Repeating the same process for $T$, one can get the 
OPE of the $T(z)T(w)$. We leave out the 
detail of the all, and just give one example of them, saying 

\begin{eqnarray}
 [\tau _{\alpha} \tau _{\beta}]_2
&&=[[\tau _{\alpha}\psi _{\beta}]_1\psi _{-\beta}]_1
+[[\tau _{\alpha}\psi _{\beta}]_2 \psi _{-\beta}]_0
+[\psi _{\beta}[\tau _{\alpha}\psi _{-\beta}]_2]_0\nonumber\\
&&=\delta _{\alpha,\beta}[\psi _{\alpha}\psi _{\beta}]_0
+(1+{\alpha}^2/k)\delta _{-\alpha,-\beta}[\psi _{\beta}\psi _{-\alpha}]_0
\nonumber\\
&&~~+\delta _{\alpha,\beta}[\psi _{\beta}\psi _{\alpha}]_0
+(1+{\alpha}^2/k)\delta _{-\alpha,\beta}[\psi _{-\alpha}\psi _{\beta}]_0
\nonumber\\
&&~~+(N_{\alpha, \beta}N_{-\alpha,\alpha+\beta}
+N_{-\alpha, \beta}N_{\alpha,-\alpha+\beta}
-\frac{(\alpha\cdot\beta)^2}{k^2})\tau _{\beta}\nonumber\\
&&~~+N_{\alpha, \beta}[[\psi _{-\alpha}\psi _{\alpha+\beta}]_0\psi _{-\beta}]_1
+N_{-\alpha, \beta} [[\psi _{\alpha}\psi _{-\alpha+\beta}]_0
\psi _{-\beta}]_1\nonumber\\
&&~~ +\frac{\alpha\beta}{k}(1+{\alpha}^2/k)
(\delta [\psi _{\alpha}\psi _{-\beta}]_0
-\delta _{-\alpha,\beta}[\psi _{-\alpha}\psi _{-\beta}]_0),
\end{eqnarray}

\noindent consider the summation for $\alpha$ and $\beta$, we have 
  
\begin{equation}
 [TT]_2=2T,
\end{equation}

\noindent the other terms can be derived in the same manner, they read 

\begin{equation}
 [TT]_4=c/2,\hskip 0.5truecm [TT]_3=0,\hskip 0.5truecm 
 [TT]_1=\partial T,
\end{equation}

\noindent equivalently, we can re-express these results as 

\begin{equation}
  T(z)T(w)=\frac{c/2}{(z-w)^4}+\frac{2T}{(z-w)^2}
+\frac{\partial T(w)}{z-w}+\ldots,
\end{equation}

\noindent in which the central charge $c$ is given by formula 
(\ref{eq:c}). The last equation is the OPE of the stress momentum 
tensor. In fact, notice that $\tau _{\alpha}=\tau _{-\alpha}$, 
and there is only single index $\al$ needs for summation, we can 
extend the summation over $\Phi$ to $\Delta$ without difficulty for 
$\tau _{\alpha}$. And if we choose the root set $\Delta$ 
for summation, no essential differences will arise, for 
$\sum _{\alpha \in \Delta}\alpha=0$. The only difference is 
the normal constant, which are two times of the original one.
So the PFs construction for Virasoro can be extended to any 
given Lie algebras. Therefore we obtain parafermion 
representation of the Virasoro algebras underlying any given  
Lie algebra with arbitrary level $k$. It is trivial to 
extend this construction to semi-simple Lie algebra case, 
if the condition is satisfied by every copy of the simple 
Lie algebra. 

\setcounter{section}{4}
\setcounter{equation}{0}
\section*{4.PFs realization of RCFT with spin $3$}

In this section we will construct a spin three field, calculate the 
OPE of the field with $T$. It turns out that the field is a primary 
field. This is one of the main features of $W_3$ algebra. 
We conjecture that this field is the first primary field in $W$ 
algebras. We then calculate the OPE of the spin $3$ field with itself, 
and in which a spin $4$ primary field emerges.    

Using the notation introduced in the previous, i.e. 
$\eta _{\beta} \equiv [\psi _{\beta}\psi _{-\beta}]_{-1}$, further we 
define $w_{\beta} \equiv \eta _{\beta}-\eta _{-\beta}$, and 
$w_3 =\sum _{\beta\in \Phi}w_{\beta}$. Obviously, $w_{\beta}=-w_{-\beta}$, 
$\sum _{\beta\in \Phi}w_{\beta}=-\sum _{\beta\in -\Phi}w_{\beta}$. 

Perform the same calculation by properly choosing of $Q,\;p,\;q$ in the 
Jacobi-like identity (the final result is the same for different choice, 
but for certain choice the calculation becomes simpler), and we have, 

\begin{equation}
 [Tw_3]_5=[Tw_3]_4=[Tw_3]_3=0,
\end{equation}

\noindent while for example, setting $Q=r=1,\;q=2$, we have

\begin{eqnarray}
 [T\eta _{\beta}]_2 &&=[\psi _{\beta}[T\psi _{-\beta}]_2]_{-1}
 +[[T\psi _{\beta}]_1\psi _{-\beta}]_0
 +[[T\psi _{\beta}]_2\psi _{\beta}]_{-1}\nonumber\\
 &&=(2-2/k)\eta _{\beta}+[\partial\psi _{\beta}\psi _{-\beta}]_0\nonumber\\
 &&=3\eta _{\beta},
\end{eqnarray}

\noindent therefore we have, 

\begin{equation}
 [Tw_{\alpha}]_2=3w_{\alpha}, \hskip 0.5truecm or, \;\;[Tw_3]_2=3w_3.
\end{equation}

\noindent In the process for deriving those results, some known identities 
are used without mention. Similarly, we obtain,

\begin{equation}
 [Tw_3]_1=\partial w_3.
\end{equation}

\noindent Equivalently, the OPE expression is 
  
\begin{equation}
 T(z)w_3(w)=\frac{3w_3(w)}{(z-w)^3}+\frac{\partial w_3(w)}{z-w}+\ldots
\end{equation}

\noindent and we complete the proof for the $w_3$ that it is a spin 
three primary field. 

However, from $\sum _{\beta\in \Phi}w_{\beta}
=-\sum _{\beta\in -\Phi}w_{\beta}$, we know that the summation of 
$w_{\alpha}$ defined on the root set $\Delta$ is identical to zero. 
It means that we cannot find any extension of the Virasoro 
algebras ($W$-algebra) on the total root 
system $\Delta$ for any Lie algebras ${\cal G}$. If we expect that such 
extension to be existence, one part of the roots (we denote it by $\Phi$), 
on which the spin three field are defined, must be separated out. 
While the restriction of the central charge require that the number of the 
elements in $\Phi$ is the half of the ones in $\Delta$, or is the same 
as in $P$. On the other hand, the consistence of the theory brings more 
constraints on the roots set $\Phi$. We can express them as 

\noindent{\it \bf Proposition:} The parafermion representation of spin three 
primary field $w_3$ is invariant under the Weyl reflection up 
to a minus one.

\begin{eqnarray}
s(w_3)=\pm w_3,\\
s_{\beta}(w_{\alpha})=w_{\alpha} 
-\frac{2(\alpha\beta)}{\beta ^2}w_{\beta},
\hskip 0.5truecm \alpha,\; \beta \in \Phi.
\end{eqnarray}

\noindent Starting from this, we have the following relation for the 
height of $\Phi$, it reads,
  
\begin{equation}
h_{\Phi}=h_{\Phi}\pm h_{\Phi},
\end{equation}

\noindent then the solution of the equation is $h_{\Phi}=0$, which 
recover the consistent condition. We used the $Z$-algebra technique 
to construct $W_3$-algebra for $SU(3)$ PFs in\cite{DFSW}, for simplicity 
we choose the symmetric roots 
$\Phi=\{\alpha _1,\alpha _2, \alpha _3=-(\alpha _1+\alpha _2)\}$ 
for summation there. At now we see that, this choice 
is essential and unique. Here, we have no enough space to 
list out the calculation in detail. The OPE of $W_3(z)W_3(w)\;\;(g>2)$ are 

\begin{eqnarray}
W_3(z)W_3(w)&&=\frac{c/3}{(z-w)^6}+\frac{2T}{(z-w)^4}
+\frac{\partial T}{(z-w)^3}
\nonumber\\
&&~~+\frac{1}{(z-w)^2}\left(2b^2\Lambda (w)+\frac{3}{10}
\partial ^2T(w)+V(w)\right)\nonumber\\
&&~~+\frac{1}{z-w}\left(b^2\partial \Lambda (w)+\frac{1}{15}
\partial ^3T(w)+\partial V(w)\right),\\ 
W_3 &&=\left(\frac{k^3}{6(k-2)(k+1)(k+g)}\right)^{1/2}w_3,
\end{eqnarray}

\noindent in which 

\begin{eqnarray}
V(z)=&&\frac{2(4k+3)k^2}{3(k-2)(k+1)(k+g)}\sum _{\alpha\in \Phi}
(\Omega _{\alpha}+\Omega _{-\alpha})\nonumber\\ 
&&-\frac{2k^2}{(k-2)(k+1)(k+g)}\sum _{\alpha,\beta\in\Phi}
\alpha\beta[\tau _{\alpha}\tau _{\beta}]_0-2b^2[TT]_0\nonumber\\
&&+\left(\frac{3}{10}(2b^2-1)-\frac{k}{2(k-2)}\right)\partial ^2T.
\label{eq:V}
\end{eqnarray}

\noindent is a spin four primary field, which can be proven from 
general CFT \cite{BPZ}, or by directly calculation. It is very 
obviously that $\sum _{\alpha,\beta\in\Delta}
\alpha\beta[\tau _{\alpha}\tau _{\beta}]_0=0$, and the left 
parts of $V$ is not a primary field anymore. So the definition of 
$V$ cannot be extended to the root set $\Delta$,  For $g=3$, 
saying the $SU(3)_k$ case, its expression reduce to\cite{DFSW}  

\begin{eqnarray}
V(z)=&&\frac{2(4k+3)k^2}{3(k-2)(k+1)(k+3)}\sum _{\alpha \in \Phi}
(\Omega _{\alpha}+\Omega _{-\alpha})\nonumber\\
&&+\frac{2k^2}{(k-2)(k+1)(k+3)}\sum _{\alpha \in \Phi}
 [\tau _{\alpha}\tau _{\alpha}]_0\nonumber\\
&&+\left(\frac{2(k+3)}{3(k-2)(k+1)}-2b^2\right)[TT]_0\nonumber\\
&&+\left(\frac{3}{10}(2b^2-1)- \frac{k}{2(k-2)}\right)\partial ^2T.
\end{eqnarray}

\noindent which is null at $k=3$ \cite{DFSW}, so the algebra 
is closed. The detailed calculation 
can be fund \cite{DFSh}. However, for $g\geq 4$ cases, the full 
solution of this problem is still open for their complexity.
In the scene that a spin $4$ primary field 
emerges, the algebra is not closed. However, for  
higher rank Lie algebra, the central charge, which reflects 
the character of symmetry is larger. From the general 
theory of CFT\cite{BPZ,Zamo},  we know that the more independent 
primary fields are needed for the larger central charge. 
From this point view, it is natural that the field is not closed at 
spin $3$. Unfortunately, we do not find an effective approach to 
calculate the number of the independent primary fields at now.  
For the well known method to enumerate the independent generating 
fields is the so called "character technique" \cite{BoSc}. 
By that technique, the independent generating fields is the same 
as the number of independent Casimirs of ${\cal G}$. 
From this point view, no spin three primary field will 
emerge in $SU(2)$ case, and this is indeed the fact from other 
approaches. But we have been obtained a $W_5$ algebra 
from the $SU(2)$ PFs. 

From the above discussion that the root set $\Phi$ forms a closed 
cycle in the root space. However, for a lot of Lie algebras, this 
condition cannot be satisfied. In fact, we know that the set $\Phi$ 
can be obtained from the positive system $P$ by some appropriate Weyl 
reflection. Obviously, this requires the height of the $P$ to be: 

\begin{equation}
h_p=2\sum _{\al \in P}n_{\alpha}, \hskip 0.5truecm 
h_{\rho}=\sum _{\al \in P}n_{\alpha}, 
\label{eq:htp}
\end{equation}

\noindent where $n_{\alpha}\in N$, is the times of the $\alpha$ 
emerging in $P$. (\ref{eq:htp}) says that, for every simple root $\al _i$, 
his times appearing in $P$ must be an even number. Obviously, one 
cannot find such set $\Phi$ for many Lie 
algebras. For $A_l$ algebra, 
the positive system is $P=\{\alpha _1, \alpha _2,\;\ldots 
{\alpha}_l,\;\alpha _1+\alpha _2,\ldots, 
\alpha _{l-1}+ \alpha _l,\;\alpha _1+\alpha _2+\alpha _3,\ldots, 
\alpha _{l-2}+\alpha _{l-1}+\alpha _l,\ldots\,\ldots,\alpha _1+\alpha _2
+\ldots+\alpha _{l-1}+\alpha _l\}$, the last one in it is the highest root. 
It is obviously that, the height of simple roots ${\al}_j$ in $P$ 
(the sum of the multiplicities of ${\al}_j$ as an element of a root in $P$) 
is $h_{\al _j}(A_l)=j(l-j+1)$. So in general, the set $\Phi$ 
does not exist for any algebras $A_{2n+1}, (n\geq 1)
(h_{\alpha _1}=2n+1,~~h_{\alpha _2}=4n,...)$, 
and $\Phi$ exists for algebras $A_{2n}$ $(n\geq 1)$ (The height 
labeled by arbitrary simple root $\al _i$ is an even number. 
For example, the height $h_{\alpha _1}=2n$ and $h_{\alpha _2}=2(2n-1)$ ). 
For $D_1=A_1$, $D_2=A_1\op A_1$, so there are no problem in these two 
cases; while for $D_3=A_3$, AND NO solution can be found in this case. 
For simplicity we leave the discussion of the algebras $D_l(\geq 4)$,  
$E_6,\;E_7,\;E_8$ and non-simple laced algebras to other place.

\setcounter{section}{5}
\setcounter{equation}{0}
\section*{5. Discussion}

In this paper, we consider a construction of $W$-symmetries (algebras) 
through a special kind of coset currents, the non-local currents 
(parafermion), which take values on the coset space $G/U(1)^r$, 
where rank $r$ Lie group $G$ is limited to a simply connected 
compact one. It turns out that, the restriction given by the 
$W$-symmetry on the PFs underlying Lie algebra is very 
stringent. Our extension is very general, all of the previously 
known results of PFs construction are very simple example of 
the present discussion. In fact, the present discussion can be 
generalized to semi-simple Lie algebra case, if the condition 
of the root set is hold for every copy of its simple Lie algebra. 

The important property of OPE is that the singular parts of it 
can be governed by certain algebra. In PFs construction we use 
part of the regular terms of the parafermionic OPE , and they satisfy 
certain algebras also. The property and relation of their higher order 
terms need further studying.   

Because of the the definition of RCFT invariant boundary state is 
directly relevant to $W$-algebra, and the important role played by 
boundary state in $Dp$-brane theory, it is reasonable to expect that 
$W$-algebra (geometry) would be found his position in the $Dp$-brane 
theory studying. The number of independent primary fields are 
relevant to fusion rule and character. So the problems need to 
further explore.

For $\psi _{\al}$ is primary currents, from the general 
discussion of the boundary state of open string, it seems 
that, a kind of new boundary state 

\beq
(\psi ^{\al} _{n} \pm (-1)^{-\al ^2/2k} \psi ^{\al} _{-n})|B>=0.
\eeq 

\no should exist. By adding appropriately chosen $U(1)$ current, 
spin $1$ currents or spin $3/2$ supercurrents, and their corresponding 
boundary states can be obtained.

\vskip 0.5cm
\no {\bf Acknowledgments:} 
One of the authors (Ding) would like to thanks H. Fan, K.J. Shi, 
Y. K. Lau, S. K. Wang and L. Zhao for fruitful discussion. 
The work was supported in part 
by the "Natural Science Foundation of China" and the "Project of 973".

\vspace{1cm}

\setcounter{section}{6}
\setcounter{equation}{0}
\section*{6. Appendix}

For the convenience of usage, here we recall some basic data 
of the root system of the Lie algebras\cite{Kac}.

Any simple Lie algebra can be classified to the classical 
series $A_l$, $B_l$, $C_l$, $D_l$, and the except ones, $E_6$, 
$E_7$, $E_8$, $F_4$ and $G_2$. There are at most two different 
length roots for any of them. For our purposes we choose the 
basis as

\begin{equation}
 [E_{\alpha}, E_{\beta}]=\varepsilon _{\alpha,\beta}E_{\alpha+\beta}
\end{equation}

\noindent Denote the root set of ${\cal G}$ by $\Delta$, We have 
$\varepsilon _{\alpha,\beta} \neq 0$, if $\alpha+\beta \in \Delta$, 
$\varepsilon _{\alpha,\beta}=0$, if $\alpha+\beta \notin \Delta$, and 

\begin{equation}
\varepsilon _{\alpha,\beta}=
-\varepsilon _{\beta,\al}=-\varepsilon _{-\alpha,-\beta}
=\frac{(\alpha+\beta)^2}{\beta ^2}\varepsilon _{-\alpha,\alpha+\beta}.
\label{eq:Con}
\end{equation}

Let $\Pi=\{\alpha _1,\;\alpha _2,\;\ldots,\alpha _r\}$ be a simple system of 
roots of ${\cal G}$, $r$ be the rank of ${\cal G}$, and Let $P$ be the 
corresponding positive system, then we have: 

(i)$\{P\}\bigcap \{-P\}=\emptyset,\; \{P\}\bigcup \{-P\}=\Delta, 
\alpha,\;\beta\in P,\; \alpha +\beta \in \Delta,\Rightarrow 
\alpha+\beta \in P$;

(ii)If, $1\leq i,\;j \leq r$, then $\alpha _i-\beta _j \notin P$, 
$\alpha _i\cdot\beta _j \leq 0$;

(iii)If $\alpha \in P, \alpha=\sum _i ^r m_i \alpha _i,\; m_i\geq 0,$ 
while $\rho$ is half the sum of the $P$, 
$\rho=\frac{1}{2}\sum _{\alpha \in P}\alpha$. Then the height of 
the root $\alpha$ is $h_{\alpha}=\sum _i ^r m_i$; 
further more we define the total height of the root system $P$,
$h_t=\sum _{\alpha\in P}  h_{\alpha}$, obviously, 
$h_t=2h_{\rho}$;

(iv)Weyl reflection
$s_{\beta}(\alpha)=\alpha -\frac{2(\alpha,\beta)}{(\beta,\beta)}
\beta \equiv \alpha -\frac{2\alpha\beta}{\beta ^2}\beta $.

If $M$ generate an irreducible representation of ${\cal G}$, 
Let $Q_m$ be the quadratic Casimir operator of the representation, 
it has an unique 
highest weight $\lambda$, then,

\begin{equation}
Q_m=\lambda (\lambda +2 \rho ).
\end{equation}
\noindent For a adjoint representation of the Lie algebra, denoting 
the accompanying quadratic Casimir operator as $Q_{\psi},\;\psi$ is the 
highest weight of the adjoint representation. Then we introduce  

\begin{equation}
g=Q_{\psi}/{\psi} ^2 =1+2 \rho\psi/ {\psi} ^2,
\end{equation}

\noindent using the data given previous, we have

\begin{equation}
 {\psi}/{\psi} ^2 =\sum _{i=1} ^{r} m_i\alpha _i/ {\alpha _i} ^2,
\end{equation}

\noindent and so, 

\begin{equation}
g=1+\sum _{i=1} ^{r} m_i=\sum _{i=0} ^{r} m_i,
\end{equation}

\noindent where $g$ is the so-called  the dual Coxeter number of the 
affine Lie algebra\cite{Kac}. More directly we can use the 
Freudenthal-de Vries Strange formula:

\begin{equation}
\frac{|\rho|^2}{g}=\frac{Dim \cal G}{12} \equiv\frac{D}{12},
\end{equation}

\noindent In the $A-D-E$ cases, $g=1+h=1+r$, where the $h$ is the height 
of the highest root.

\bebb{ABC}

\bbit{Arg}
P. C. Argres and K.R. Dienes, \plb{387}{727}{1996}

\bibitem{BBSS}
F.Bais, P. Bouwknegt, M. Surridge and K. Schoutens, Nucl. Phys. {\bf B304}, 
348; 371(1988). 

\bibitem{BPZ}
A.A. Belavin. A. M. Polyakov, A. B. Zamolodchikov, 
Nucl. Phys. {\bf B241}, 333(1984).

\bibitem{BoSc}
P. Bouwknegt and K. Schoutens, Phys. Rep. {\bf 223}, 183(1993).

\bbit{Card}
J. L. Cardy, \np{324}{581}{1989}.

\bbit{CES}
H. Chaker, A. Elfallah and E. H. Saichi, \cqg{14}{2049}{(1997)}.

\bibitem{DFSh}
X.M. Ding, H. Fan, K.J. Shi, and P. Wang, Commun. Theor. Phys. 
{\bf23}, 197(1996).

\bibitem{DFSW}
X.M. Ding, H. Fan, K.J. Shi, P. Wang and C.Y. Zhu, Phys. Rev. Lett. 
{\bf 70}, 2228(1993); Nucl. Phys. {\bf B422},  307(1994). 

\bibitem{DiQi}
J. Distler and Z. Qiu, Nucl. Phys. {\bf B336}, 533(1990).

\bibitem{FaLy}
V.A. Fateev and S.L. Lykyanov, Int. J. Mod. Phys. {\bf A3}, 507(1988).

\bibitem{FaZa}
V.A.Fateev and A.B.Zamolodchikov, Nucl.Phys. {\bf B280}, 644(1987),

\bibitem{FORT}
L. Feher, L.O'Raifeartaich, P. Ruelle, I. Tsutsui and A. Wipf, 
Phys. Rep. {\bf 222}, 1(1992). 

\bbit{Gepn}
D. Gepner, Nucl. Phys. {\bf B290}, 10(1987).

\bibitem{Gep}
D. Gepner, Nucl. Phys. {\bf B296}, 757(1988);  
Phys. Lett. {\bf B 199}, 380(1987)

\bibitem{GeQi}
D.Gepner and Z.Qiu, Nucl. Phys.{\bf B285}, 423(1987).

\bbit{GKS}
A. Giveon, D. Kutasov and N. Seiberg, 
{\it Adv. Theor. Math. Phys.} {\bf 2}, 733(1998); hep-th/9806194.

\bbit{Green}
M.B.Green, \plb{201}{42}{1987}.

\bbit{GKP}
S.S. Gubser, I.R. Klebanov and A. M. Polyakov, \plb{428}{105}{1998}; 
hep-th/9802109.

\bbit{GSa}
M. Gutperle and Y.Satoh, Nucl. Phys. {\bf B543}, 73(1999), hep-th/9808080; 
Nucl.Phys. {\bf B555}, 477(1999), hep-th/9902120.

\bbit{Hull}
C.M.Hull, {\it Lectures on $W$-Gravity, $W$-Geometry and $W$-String}, 
hep-th-9302110.

\bbit{Ish}
N. Ishibashi, {\it Mod. phys. Lett.}{\bf A4}, 251(1989).  

\bibitem{Kac}
V. G. Kac, {\it Infinite Dimensional Lie Algebras}, Cambridge University, 
1990. 

\bbit{KS}
D. Kutasov and N. Seiberg, {\it JHEP} {\bf 9904} 008(1999); hep-th/9903219.

\bbit{Mad}
J. Maldacena, {\it Adv. Theor. Math. Phys.} {\bf 2}, 231(1998); 
hep-th/9711200.

\bbit{Maro}
V. Maroto, \np{527}{717}{(1998)}.

\bibitem{Neme}
D. Nemeschansky, Phys. Lett. {\bf B 224}, 121(1989).

\bbit{PCai}
J. Polchinski and Y. Cai, \np{296}{91}{1988}.

\bbit{RSc}
A. Recknagel and V. Schomerus, Nucl. Phys. {\bf B531}, 185(1998); 
hep-th/9712186.

\bibitem{Waki}
M. Wakimoto, Commun. Math. Phys. {\bf 104}, 605(1986).

\bibitem{WaDi}
P. Wang and X. M. Ding, Phys.Lett. {\bf B335}, 56(1994). 

\bibitem{West}
For a review see, P. West, A review of $W$ strings, preprint 
Goteborg-ITP-93-40.  

\bbit{Witt}
E.Witten, {\it Adv. Theor. Math. Phys.} {\bf 2}, 253(1998); 
hep-th/9902150.

\bibitem{Yang}
S. K. Yang, Nucl. Phys. {\bf B285}, 639(1987). 

\bbit{YZh}
M. Yu and B. Zhang, Nucl.Phys. {\bf B551}, 425(1999); 
hep-th/9812216. 

\bibitem{Zamo}
A.B. Zamolodchikov, Theor. Math. Phys. {\bf 65}, 1205(1985).

\bibitem{ZaFa}
A.B.Zamolodchikov and V.A.Fateev, Sov. Phys, JETP {\bf 62}, 215(1985).

\eebb

\end{document}